\documentclass[aps,pra,twocolumn,superscriptaddress,showpacs]{revtex4}
\usepackage{amsmath}
\usepackage{graphicx}

\newcommand{\ket}[1]{\mbox{$\mid \! #1 \, \rangle$}}
\newcommand{\bra}[1]{\mbox{$\langle \, #1 \! \mid$}}

\usepackage{epsfig}

\begin{document}

\title{Entanglement and communication-reducing properties of noisy $N$-qubit states}

\author{Wies{\l}aw Laskowski}
\affiliation{Institute of Theoretical Physics and Astrophysics, University of Gda\'nsk, PL-80-952 Gda\'nsk, Poland}
\affiliation{Fakult\"at f\"ur Physik, Ludwig-Maximilians Universit\"at M\"unchen, D-80799 M\"unchen, Germany}
\affiliation{Max-Planck Institut f\"ur Quantenoptik, D-85748 Garching, Germany}

\author{Tomasz Paterek} 
\affiliation{Institute for Quantum Optics and Quantum Information, Austrian Academy of Sciences, A-1090 Vienna, Austria}
\affiliation{Centre for Quantum Technologies and Department of Physics, National University of Singapore, 117542 Singapore}

\author{{\v C}aslav Brukner}
\affiliation{Institute for Quantum Optics and Quantum Information, Austrian Academy of Sciences, A-1090 Vienna, Austria}
\affiliation{Faculty of Physics, University of Vienna, A-1090 Vienna, Austria}

\author{Marek \.Zukowski}
\affiliation{Institute of Theoretical Physics and Astrophysics, University of Gda\'nsk, PL-80-952 Gda\'nsk, Poland}

\begin{abstract}
We consider properties of states of many qubits, which  arise after sending certain entangled states via various noisy channels (white noise, coloured noise, local depolarization, dephasing and amplitude damping). Entanglement of these states is studied and their ability  to violate certain classes of Bell inequalities.
States which violate them allow for higher than classical efficiency of solving related distributed computational tasks with constrained communication. This is a direct property of such states -- not requiring their further 
modification via stochastic local operations and classical communication  such as entanglement purification or distillation procedures. We identify novel families of multi-particle states which are entangled but nevertheless allow local realistic description of specific Bell experiments.
For some of them, the ``gap'' between the critical values for entanglement and violation of Bell inequality remains finite even in the limit of infinitely many qubits.
\end{abstract}


\date{\today}

\maketitle

\section{Introduction}

Despite a considerable progress in understadning entanglement the 
question whether every entangled state
does not admit a local realistic simulation 
is as yet unanswered.
Bell has shown that certain pure entangled states 
violate constraints imposed by local hidden variable models~\cite{BELL}.
Bell's result was generalized by Gisin and Peres who demonstrated the violation for all 
bipartite pure entangled states~\cite{GISIN,GISINPERES}.
Popescu and Rohrlich
showed that no local realistic description
is possible for any pure multipartite entangled state; the proof involved post-selection~\cite{POPESCU}.
Without post-selection, it is not clear whether there are pure entangled states
which admit local realistic model for all possible measurements. 
Bell experiments with two settings per observer
in which only correlation functions are measured
indeed admit local hidden-variable explanation
even for some pure entangled states \cite{GEN_GHZ1,GEN_GHZ2}.

For mixed states, this relation is even subtler. 
Werner states are an example of bipartite entangled mixed states which allow a local realistic model for direct measurements \cite{WERNER,BARRETT}.
Almeida {\it et al.} found recently that the range of entanglement admixture for which the state of two $d$-level systems is
both entangled and admits local hidden-variable model for all measurements decreases proportionally to $\log(d)/d$~\cite{ALMEIDA}.
Also some genuinely tripartite entangled mixed states can admit hidden variable description
for all measurements \cite{TA}. 
It was shown that entangled states
upon sequential local measurements  may be transformed into ones that do not allow a local realistic description \cite{HN1,HN2,HN3,HN4}.
Note, however, that this is not a {\em direct} property of such states, only the final states which result out of such transformations are endowed with it. 
The relation between entanglement and local realism
for multipartite mixed states is still largely unexplored.
Our work addresses this problem.

This relation is not only of importance for fundamental research, 
but also in the context of quantum communication and quantum computation. For
certain tasks, such as quantum communication complexity problems~\cite{QCCP1,QCCP2}
or device-independent quantum
key distribution~\cite{KEY_DIST1,KEY_DIST2}, entangled states are useful only to the extent
that they violate Bell inequalities. 
Furthermore, entangled
states which violate certain Bell inequalities, but satisfy other ones,
are useful for particular quantum communication complexity problems
directly related with the violated inequalities (for details of the link between the inequalities and communication complexity problems see \cite{QCCP2}).

In such problems, several partners have disjoint sets of data, and under a strict communication constraint, to e.g. one bit per partner, are asked to give the value of a task function which depends on all data.
The amount of violation of a Bell inequality for correlation functions related to the problem is proportional to the increase of the probability to get the correct value of  the task function, which quantum protocols involving the entangled states allow in  comparison with the optimal classical protocol.
Note that often additional post-processing of experimental data requires additional classical communication
and therefore increases communication complexity of quantum protocols.
Therefore, states which violate certain Bell inequalities after sequential measurements or post-selection
are usually less efficient in terms of communication complexity reduction than the states which violate the inequalities directly.
It is thus important to make both classifications of entangled states:
into admitting and not admitting local realistic models,
and into violating and not violating a given Bell inequality.

Entanglement and Bell violation of different noisy states
has already been studied by several authors \cite{ANTIA2005,SU,JCKL2006,LEANDRO,MAURO}.
All this indicates that entanglement and impossibility of a local hidden variable model 
are not only different concepts, but also truly different resources \cite{SCARANI2}.
Our aim is to identify a class of states that demonstrates this difference in a striking way.

We consider states of $N$ two-level systems resulting from sending different entangled states via noisy channels.
Noisy states are of special importance
as they take into account errors inevitable in any laboratory.
More specifically, we study white noise admixture which is often used to model imperfections of setups involving
single crystal in which spontaneous parametric down-conversion takes place.
We consider also colored noise admixture which was shown to be appropriate, e.g., in description of states generated in multiple entanglement swapping \cite{ADITI}.
Typical noisy channels (depolarization, dephasing, amplitude damping)
which act independently on every qubit are also studied.
They find applications in modeling random environment and dissipative processes \cite{NIELSENCHUANG}.

Here we find states for which even an infinitesimal small admixture
of infinitesimal weak entangled state results in a non-separable state,
while to violate standard Bell inequalities (with two settings per party) the admixture has to scale at least as $1/\sqrt{d}$,
where $d$ is the dimension of $N$ qubits, i.e. $d=2^N$. This shows a remarkable ``gap'' between the critical parameters for entanglement and for violation of standard Bell's inequalities. 
We observe that keeping the same amount of noise,
but changing the type of noise drastically changes entanglement and communication-reducing properties of the states,
i.e., whether the states allow for higher than classical reduction of communication complexity.
Furthermore, we find mixed states for which this gap remains finite even in the limit of infinitely many qubits.

\section{Toolbox}
Our tools consist of entanglement criterion in terms of correlation functions~\cite{BADZIAG},
which will prove handy for comparison with conditions for violation of Bell inequalities.
We shall take into account sets of Bell inequalities for two and more measurement settings \cite{WZ,WW,ZB,WUZONG1,WUZONG2,LPZB}.
We now describe these tools in more detail.

Arbitrary state of many qubits can be decomposed into:
\begin{equation}
\rho = \frac{1}{2^N} \sum_{\mu_1,...,\mu_N=0}^3 T_{\mu_1...\mu_N}
\sigma_{\mu_1} \otimes ... \otimes \sigma_{\mu_N},
\label{STATE}
\end{equation}
where $\sigma_{\mu_n} \in \{\openone,\sigma_x,\sigma_y,\sigma_z\}$ is the
$\mu_n$th local Pauli operator of the $n$th party ($\sigma_0= \openone$) and $T_{\mu_1...\mu_N} \in [-1,1]$ are the
components of the (real) extended correlation tensor $\hat T$. 
They are the expectation values 
$T_{\mu_1...\mu_N} = \mbox{Tr}[\rho (\sigma_{\mu_1} \otimes ... \otimes
\sigma_{\mu_N})]$.
Thus, description in terms of correlation tensor is equivalent to description in terms of density operator.
Fully separable states are endowed with fully separable extended correlation tensor,
$\hat T^{\mathrm{sep}} = \sum_{i} p_i \hat T^{\mathrm{prod}}_i$,
where  $\hat T^{\mathrm{prod}}_i = \hat T^{(1)}_i \otimes ... \otimes \hat T^{(N)}_i$
and each $\hat T^{(n)}_i$ describes a pure one-qubit state.
A state $\rho$, with correlation tensor $\hat T$, is entangled if there exists a $G$ such that \cite{BADZIAG}:
\begin{equation}
\max_{\hat T^{\mathrm{prod}}} (\hat T, \hat T^{\mathrm{prod}})_G <  (\hat T, \hat T)_G = ||\hat T||^2_{G},
\label{CRITERION}
\end{equation}
where maximization is taken over all product states
and $(.,.)_G$ denotes a generalized scalar product, with a positive semidefinite metric $G$.
We focus on diagonal $G$'s,
for which the scalar product is given by
\begin{equation}
(\hat T, \hat T')_G  = \sum_{\mu_1,...,\mu_N=0}^3 T_{\mu_1...\mu_N} G_{\mu_1...\mu_N} T'_{\mu_1...\mu_N}.
\label{GEN-SCALAR}
\end{equation}
The criterion is valid also when the sums of (\ref{GEN-SCALAR}) run through the values $j_n = 1,2,3$,
which will be often referred to as $x,y,z$.

We compare this entanglement criterion with criteria for violation of Bell inequalities.
It was shown that a simple sufficient condition for
existence of a local realistic description of the correlation
functions obtained in any Bell experiment with two measurement settings per
observer has the following form \cite{ZB}:
\begin{equation}
\mathcal{C} \equiv \max \sum_{j_1,...,j_N = 1}^2 T_{j_1...j_N}^2 \le 1,
\label{ZBCOND}
\end{equation}
where maximization is taken over all possible independent choices of
local planes in which the two settings lie. 
This condition is necessary and sufficient in the case of two qubits \cite{HORODECKIS_BELL_NS}.

We shall also use another necessary and sufficient condition,
this time for violation of a set of tight Bell inequalities with many
measurement settings per observer \cite{LPZB}.
For the case of $N+1$ observers,
all of which but the last one choose between four settings,
and the last one between two settings, this condition reads
\begin{equation}
\mathcal{D} \equiv \max \sum_{j_1(k),...,j_N(k)=1}^{2} \sum_{k=1}^2 T_{j_1(k)...j_{N}(k)k}^2 \le 1,
\label{ZBCOND1}
\end{equation}
where maximization is over all possible independent choices of local Cartesian frame basis vectors used by the observers to fix the measurement directions determining the correlation tensor components. That is, we allow each observer to define his/her 
triad of orthogonal basis directions, which define the correlation tensor components.
This condition is more demanding than~(\ref{ZBCOND})
because the coordinate systems denoted by the indices
$j_1(1)$, ..., $j_N(1)$ do not have to be the same 
as $j_1(2)$, ..., $j_N(2).$
It is necessary for the existence of
local realistic model \cite{LPZB}, or equivalently, its violation is sufficient for non existence of such models.

\section{Noises}

The states to be studied here are of two general types:
(i) Mixtures of an entangled state $\rho$
and white or colored noise $\rho_{\mathrm{noise}}$:
\begin{equation}
\rho(\Upsilon) = \Upsilon \rho + (1-\Upsilon) \rho_{\mathrm{noise}},
\label{NOISY_STATE}
\end{equation}
where $\Upsilon$ is the entanglement admixture.
(ii) States arising from local noisy channels \cite{NIELSENCHUANG}, i.e. of the form
\begin{equation}
(\mathcal{E} \otimes ... \otimes \mathcal{E})(\rho) = \frac{1}{2^N} \!\!\!\! \sum_{\mu_1,...,\mu_N=0}^3 \!\!\!\! T_{\mu_1...\mu_N} \mathcal{E}(\sigma_{\mu_1}) \otimes ... \otimes \mathcal{E}(\sigma_{\mu_N}),
\label{LOCAL_NOISE}
\end{equation}
where $\mathcal{E}$ is a map describing depolarization, dephasing or amplitude damping of a single qubit.
According to (\ref{LOCAL_NOISE}), such noises are fully described by their action on local Pauli operators.

Each type of noise considered is parameterized with a single variable
(it is either entanglement admixture $\Upsilon$ or the strength of local decoherence)
and therefore the resulting states are also characterized by this variable.
We choose the parameters such that value `$1$'
corresponds to no noise, whereas value `$0$' corresponds
to total noise which immediately destroys all initial entanglement.
Using the described separability criterion
we determine 
threshold value of the parameter,
above which the resulting state 
is entangled.
Next, using the conditions (\ref{ZBCOND}) and (\ref{ZBCOND1}) for the described Bell inequalities
we find maximal parameter
below which the state does not violate them.
Finally, we contrast these two critical values.
We summarize our results in Table \ref{TAB_SUMMARY}.

\subsection{White noise}
``White noise'' is represented by a totally mixed state
$\rho_{\mathrm{noise}} = \tfrac{1}{2^N} \openone$,
where $N$ gives the number of qubits.
A channel introducing the white noise 
to a system is the globally depolarizing channel:
\begin{equation}
\mathcal{E}_{\Upsilon}(\rho) = \Upsilon \rho + (1-\Upsilon) \tfrac{1}{2^N} \openone.
\end{equation}
Therefore, the correlation tensor of the state after the globally
depolarizing channel, $\hat T'$, is related
to the initial state by the admixture parameter $\hat T' = \Upsilon \hat T$.
The operator-sum representation
\begin{equation}
\mathcal{E}_{\Upsilon}(\rho) = \Upsilon \rho + \frac{1-\Upsilon}{2^{2N}} \sum_{\mu_1, ..., \mu_N = 0}^{3} \!\!\!\! \sigma_{\mu_1} \otimes ... \otimes \sigma_{\mu_N} \rho \sigma_{\mu_1} \otimes ... \otimes \sigma_{\mu_N},
\end{equation}
reveals that white noise admixture acts in a correlated way on all the qubits.

\subsection{Colored noise}

We will consider colored noise represented by a product state
$\rho_{\mathrm{noise}} = | 0 \rangle \langle 0 | \otimes ... \otimes | 0 \rangle \langle 0 |$,
where $| 0 \rangle$ is the eigenstate of local $\sigma_z$ Pauli operator.
Such a noise brings perfect correlations along $z$ directions to the system.

\subsection{Local depolarization}

In many cases, noise affects independently every qubit.
For example, local depolarization
can be caused by a random environment
acting autonomously on each subsystem.
The local depolarization is defined
for a single qubit in the familiar way
\begin{equation}
\mathcal{E}_p(\rho) = p \rho + (1-p) \tfrac{1}{2} \openone,
\end{equation}
i.e. it mixes the local state with the white noise
where $p$ describes the fraction of initial state still present after the decoherence.
To see the effect of local depolarization on many qubits
we find its effect on local Pauli operators
\begin{eqnarray}
\mathcal{E}_p(\openone) = \openone, & \qquad &
\mathcal{E}_p(\sigma_x) = p \sigma_x, \nonumber \\
\mathcal{E}_p(\sigma_y) = p \sigma_y, & \qquad &
\mathcal{E}_p(\sigma_z) = p \sigma_z,
\end{eqnarray}
and follow formula (\ref{LOCAL_NOISE}).

In general, the critical values arising from local depolarization
and white noise can be different.
However, in our case the critical parameters turn out to be the same
due to the structure of violation conditions for the Bell inequalities
and the form of entanglement criterion we use.
Since these conditions involve only $N$-party correlations,
local depolarization introduces a factor of $p^N$ to the elements of correlation tensor entering them
while white noise admixed to the system introduces a factor of $\Upsilon$.
Therefore, the critical values obtained using $p^N$ and $\Upsilon$ are equal,
$p_{\mathrm{cr}}^N = \Upsilon_{\mathrm{cr}}$,
independently of the state for which they are computed
(the numerical value can of course vary from state to state).

\subsection{Dephasing}

Local depolarization describes gradual
loss of coherence in all bases.
It may happen that coherence is lost in a preferred basis.
This type of noise is described by a dephasing channel
defined by its action on local Pauli operators:
\begin{eqnarray}
\mathcal{E}_{\lambda}(\openone) = \openone, & \qquad &
\mathcal{E}_{\lambda}(\sigma_x) = \sqrt{\lambda} \sigma_x, \nonumber \\
\mathcal{E}_{\lambda}(\sigma_y) = \sqrt{\lambda} \sigma_y, & \qquad &
\mathcal{E}_{\lambda}(\sigma_z) = \sigma_z,
\end{eqnarray}
where the $\sigma_z$ basis is chosen to be preferred by decoherence,
and $\lambda$ describes its strength.
Clearly, for $\lambda = 1$ the initial state is unchanged
and for $\lambda = 0$ the final state has only classical correlations
along local $z$ directions.

\subsection{Amplitude damping}

Amplitude damping channel is used to describe
energy dissipation from a quantum system.
Under amplitude damping, a system has a finite
probability, $\gamma$, to loose an excitation.
In terms of local Pauli operators this channel is descried as
\begin{eqnarray}
\mathcal{E}_{\gamma}(\openone) = \openone + (1-\gamma) \sigma_z, & \qquad &
\mathcal{E}_{\gamma}(\sigma_x) = \sqrt{\gamma} \sigma_x, \nonumber \\
\mathcal{E}_{\gamma}(\sigma_y) = \sqrt{\gamma} \sigma_y, & \qquad &
\mathcal{E}_{\gamma}(\sigma_z) = \gamma \sigma_z.
\end{eqnarray}
Note that the components of the correlation tensor
of a state after amplitude damping which contain the $z$ indices
are given by the sums of initial correlation tensor components
with both $z$ indices and zero indices, e.g.
\begin{equation}
T_{\underbrace{z...z}_{k}0...0}' = \!\!\!\! \sum_{l_1,..,l_k = \{0,3\}} \!\!\!\! \!\!\!\! T_{l_1...l_k0...0} (1-\gamma)^{n_0} \gamma^{n_3},
\end{equation}
where $n_{0} \equiv \sum_{j=1}^k \delta_{l_j,0}$ gives the number of indices $l_1,\dots, l_k$ equal to $0$,
and similarly $n_{3} \equiv \sum_{j=1}^k \delta_{l_j,3} = k-n_0$
denotes the number of indices equal to $3$.

Having described the noises of interest,
we move to studies of their influence
on certain classes of initially entangled states.

\section{Noisy states} 

We begin with the Bell state of two qubits
and mix it with white and colored noise.
The state with white noise is the Werner state
known to admit local hidden variable model for certain admixtures despite being entangled.
Interestingly, the states with colored noise,
which are maximally entangled mixed states \cite{MEMS1, MEMS2},
will be shown to be entangled and not to violate
standard Bell inequalities in an even bigger range of mixing.
We then show similar results for the GHZ states and generalized GHZ states.
For some of them, the critical admixture of entanglement
below which the state admits local hidden variable model
scales polynomially with dimension of the system,
and the mixed state is entangled already for infinitesimally small
admixture of infinitesimal entanglement.

Next, we discuss noisy states arising from independent local decoherence.
We start with generalized GHZ states as initial states
and show that even in the limit of infinitely many qubits
there is still a finite gap between critical parameter for entanglement
and the one for violation of Bell inequalities.
We show similar results when the initial state is a W state.
Roughly speaking, a simple application of the entanglement criterion detects
entanglement at least quadratically better than the Bell inequalities, i.e. the critical value for entanglement 
is at most equal to the square of the critical value to satisfy the Bell inequalities.

\subsection{Bell state}

\subsubsection{White noise}

We first rederive known results for a Werner state of two qubits with our tools.
It is a mixture of a maximally entangled state $\rho = |\phi^+ \rangle \langle \phi^+|$
and white noise $\rho_{\mathrm{noise}} = \frac{1}{4} \openone$,
where $|\phi^+ \rangle = \frac{1}{\sqrt{2}}(\ket{00} + \ket{11})$ 
and $\ket{0}$ ($\ket{1}$) denotes the eigenstate of $\sigma_z$ operator
with eigenvalue $+1$ ($-1$).
The family of Werner states is an archetypical example of a state set which
contains states that do not violate Bell inequalities despite being entangled.

Since the white noise state exhibits no correlations, the
correlation tensor components $T_{j_1j_2}'$ of the Werner state
are related to the components $T_{j_1j_2}$ of $|\phi^+ \rangle$ by the
admixture factor, $T_{j_1j_2}' = \Upsilon T_{j_1j_2}$.
The only non-vanishing correlation tensor elements
of maximally entangled states lie on the diagonal and are equal to $\pm 1$ 
(this is so when the two-particle correlation tensor is put in a Schmidt form).
If one chooses to sum over $j_n = 1,2,3$ in the scalar products of criterion (\ref{CRITERION})
the left-hand side is given by the maximal Schmidt
coefficient of the correlation tensor. For the Werner state it equals $\Upsilon$.
The right-hand side reads $3 \Upsilon^2$.
Thus, the criterion reveals entanglement for all the states of the family,
i.e., for $\Upsilon_{\mathrm{ent}} > \frac{1}{3}$.
On the other hand, the necessary and sufficient condition
for local realistic model, in the case of a standard two-settings-per-partner Bell experiment (\ref{ZBCOND}), is satisfied for
$\Upsilon_{\mathrm{lr}} \le \frac{1}{\sqrt{2}}$.
Thus, for a considerable range of $\Upsilon \in (\frac{1}{3},\frac{1}{\sqrt{2}}]$
the state is entangled, nevertheless Bell experiments involving standard inequalities have a local realistic explanation.
One could call this range of $\Upsilon$ a ``Werner gap''.

\subsubsection{Colored noise}

Interestingly, changing the type of noise from white to colored
influences both entanglement of the state and possibility of local realistic model.
We have investigated critical admixtures of different types of noise,
above which condition (\ref{ZBCOND}) is satisfied
and summarize them in Table \ref{phi}.
Changing the type of colored noise alone,
although does not change entanglement threshold of the state,
dramatically influences its communication-reducing properties.
The splitting of this table into different 
rows  is motivated by different relations
between correlations present in the noise and in the entangled state $| \phi^+ \rangle$.
In the first row, the white noise has no correlations.
In the second row, the noises have some of the correlations of the $| \phi^+ \rangle$ state.
Therefore, for all $\Upsilon > 0$ there are perfect correlations in the system (in the basis of states of noise)
and additionally at least some correlations in a complementary measurement directions.
This explains the violation of a two-setting Bell inequality \cite{ESSENCE}.
In the third row, the noise has exactly opposite correlations
to those present in the $| \phi^+ \rangle$ state.
In the last row, the noises have  correlations of a different character than those of the entangled state.

\begin{table}
\begin{center}
\begin{tabular}{c | c | c}
\hline  \hline
Type of noise & Entanglement & Comm. reduction  \\  \hline \hline
$\openone \otimes \openone$ & $\Upsilon> \frac{1}{3}$& $\Upsilon>\frac{1}{\sqrt{2}} = 0.70711$\\ \hline
$\ket{\pm}{}_z {}_z\bra{\pm} \otimes \ket{\pm}{}_z {}_z\bra{\pm} $ & $\Upsilon>0$ & $\Upsilon>0$ \\
$\ket{\pm}{}_y {}_y\bra{\pm} \otimes \ket{\mp}{}_y {}_y\bra{\mp} $ & & \\ 
$\ket{\pm}{}_x {}_x\bra{\pm} \otimes \ket{\pm}{}_x {}_x\bra{\pm} $ &  &  \\ \hline
$\ket{\pm}{}_z {}_z\bra{\pm} \otimes \ket{\mp}{}_z {}_z\bra{\mp} $ & $\Upsilon>0$ & $\Upsilon>\frac{1}{\sqrt{2}} = 0.70711$ \\
$\ket{\pm}{}_y {}_y\bra{\pm} \otimes \ket{\pm}{}_y {}_y\bra{\pm} $ &  &  \\
$\ket{\pm}{}_x {}_x\bra{\pm} \otimes \ket{\mp}{}_x {}_x\bra{\mp} $ & &  \\ \hline
$\ket{\pm}{}_x {}_x\bra{\pm} \otimes \ket{\pm}{}_y {}_y\bra{\pm} $ & $\Upsilon>0$ & $\Upsilon>0.56731$ \\
$\ket{\pm}{}_x {}_x\bra{\pm} \otimes \ket{\mp}{}_y {}_y\bra{\mp} $ &  &  \\
$\ket{\pm}{}_x {}_x\bra{\pm} \otimes \ket{\pm}{}_z {}_z\bra{\pm} $ &  &  \\
$\ket{\pm}{}_x {}_x\bra{\pm} \otimes \ket{\mp}{}_z {}_z\bra{\mp} $ &  &  \\
$\ket{\pm}{}_y {}_y\bra{\pm} \otimes \ket{\pm}{}_z {}_z\bra{\pm} $ &  &  \\
$\ket{\pm}{}_y {}_y\bra{\pm} \otimes \ket{\mp}{}_z {}_z\bra{\mp} $ &  &  \\
\hline \hline
\end{tabular}
\end{center}
\caption{The table presents critical value of entanglement admixture
above which the
two-qubit state $\Upsilon | \phi_+ \rangle \langle \phi_+ | + (1-\Upsilon) \rho_{\rm{noise}}$
is entangled (middle column) and allows reduction of communication complexity, 
i.e. violates standard Bell inequalities (right column),
for different types of noise (left column). 
In the left column, $\openone \otimes \openone$ denotes the white noise
and e.g. $\ket{\pm}{}_k {}_k\bra{\pm} \otimes \ket{\mp}{}_l {}_l\bra{\mp}$
denotes the colored noise which is a product state of either $\ket{+}{}_k {}_k\bra{+} \otimes \ket{-}{}_l {}_l\bra{-}$
or $\ket{-}{}_k {}_k\bra{-} \otimes \ket{+}{}_l {}_l\bra{+}$, where $\ket{\pm}{}_k$
is the eigenstate of Pauli $\sigma_k$ operator with eigenvalue $\pm 1$
(either the upper signs enter the states of the noise or the lower signs).}
\label{phi} 
\end{table}

The Werner states are not the ones with the largest possible gap. 
For example, if one admixes, e.g., colored noise $\rho_{\mathrm{noise}} = | \pm \rangle_z {}_z\langle \pm | \otimes | \mp \rangle_z {}_z \langle \mp |$ to the $|\phi^+ \rangle$ Bell state, the resulting state is entangled already for
an infinitesimally small value of $\Upsilon$,
and it satisfies condition (\ref{ZBCOND}) for all $\Upsilon_{\mathrm{lr}} \le \frac{1}{\sqrt{2}}$.
Thus, the range of $\Upsilon$ for which the state admits local realistic model for standard correlation Bell experiments
and is still entangled
is much larger than for the Werner state.
Moreover, this is the maximal possible range (there is no other state and dichotomic measurements which would give bigger Werner gap)
because the critical value $\Upsilon_{\mathrm{lr}} = \frac{1}{\sqrt{2}}$
corresponds to the maximal violation of local realism \cite{TSIRELSON}.
We note that such mixed states are known to be maximally entangled \cite{MEMS1,MEMS2}.

\subsection{GHZ state}

\subsubsection{White noise}

The presented tools allow us
to construct and investigate entangled states of multiple qubits, with a non-zero Werner gap, in a systematic way. 
We first consider the Werner-like states of $N$ qubits which are mixtures of
the GHZ state $| \mathrm{GHZ}_N \rangle = \frac{1}{\sqrt{2}}(\ket{0 \dots 0} + \ket{1 \dots 1})$ and the white noise. 
Using criterion (\ref{CRITERION}) one finds $\Upsilon_{\mathrm{ent}} = 1/(2^{N-1}
+ 1)$ for the critical admixture above which the state is entangled \cite{BADZIAG,PR}. 
The critical value for violation of a complete set of standard Bell
inequalities for correlation functions equals 
$\Upsilon_{\mathrm{lr}} = 1/\sqrt{2^{N-1}}$ (see \cite{ZB}).
Therefore, for $\Upsilon \in (\frac{1}{2^{N-1}+1},\frac{1}{\sqrt{2^{N-1}}}]$
the state is entangled but all two-setting correlation Bell experiments
admit local realistic model.
Also multisetting inequalities of Ref. \cite{LPZB} are all satisfied in this range.

To illustrate how the range of the Werner gap can depend on the Bell inequality,
we consider inequalities of Refs. \cite{ZUK,NLP}.
If one considers all possible settings, restricted to one measurement plane on the Bloch sphere for each observer,
the critical value for violation of local realism
changes to $\Upsilon_{\mathrm{lr}}^{\infty} = 2 (2/\pi)^N$, see \cite{ZUK}, and therefore decreases the Werner gap.
This result  is a limiting case
for inequalities involving $M$ settings per party studied in \cite{NLP}. These inequalities involve measurement settings (again in a specific plane for each observer) evenly spaced at the Bloch sphere. 
One has  $\Upsilon_{\mathrm{lr}}^{\infty} = \lim_{M \to \infty} \Upsilon_{\mathrm{lr}}^{M}$ \cite{NLP}, notation is obvious here.
One may ask for how many settings the critical entanglement admixture for violation of local realism
for finite and continuum number of settings are already very close.
For bigger $M$, one finds using Taylor series that $\Upsilon_{\mathrm{lr}}^{M} = \Upsilon_{\mathrm{lr}}^{\infty}[1 + \frac{\pi^2}{24} \frac{N-3}{M^2} + \mathrm{O}(\frac{N^2}{M^4})]$.
If one neglects all the small terms of $\mathrm{O}(\frac{N^2}{M^4})$,
the relative error $\epsilon = (\Upsilon_{\mathrm{lr}}^{M}-\Upsilon_{\mathrm{lr}}^{\infty})/\Upsilon_{\mathrm{lr}}^{\infty}$
is given by $\epsilon \approx 4 \pi^2 \frac{N-3}{M^2}\%$.
Thus, for $M = N$ the two critical admixtures are close even for a few particles ($\epsilon$ smaller than $4\%$ for all $N \ge 4$).

\subsubsection{Colored noise}

Similarly to the case of the Bell states,
also for the GHZ state the range of the Werner gap
depends on the type of admixed noise.
For the odd-$N$ GHZ states the correlations $T_{z...z}$  vanish,
and it is interesting to consider the colored noise $\rho_{\mathrm{noise}} = | 0 \rangle \langle 0 |^{\otimes N}$
which re-introduces the missing correlations.
The full correlation tensor of $\rho(\Upsilon)$, i.e., the one covering ``Greek'' indices from $0$ to $3$, has the following
non-vanishing components: $T_{z...z} = 1- \Upsilon$,  and also $2^{N-1}$
components with $2k$ indices equal to $y$ and the remaining indices
equal to $x$ (where $k = 0,1,...,\frac{N-1}{2}$). These latter ones  are given by $(-1)^k \Upsilon$. 
Finally, one finds $2^{N-1}-1$ components with $2k$ indices (where $k = 1,...,\frac{N-1}{2}$) set at $0$ and the remaining indices
set to $z$. All these have the value of $1$. 
Consider a metric $G$ with only the following nonzero elements: $G_{zz0...0} = \Upsilon$
and $G_{i_1...i_N}=1$ for the components with $2k$ indices equal to $y$ and the rest equal to $x$.
For such a metric, the maximum of the scalar product on the left-hand side of condition (\ref{CRITERION}) is equal to $\Upsilon$.
The right-hand side equals $||\hat T||_G^2 = \Upsilon + 2^{N-1} \Upsilon^2$, which is always greater than $\Upsilon$.
Thus, the state is entangled already for an infinitesimally small $\Upsilon$.

\begin{figure}
\begin{center}
\includegraphics[scale=0.6]{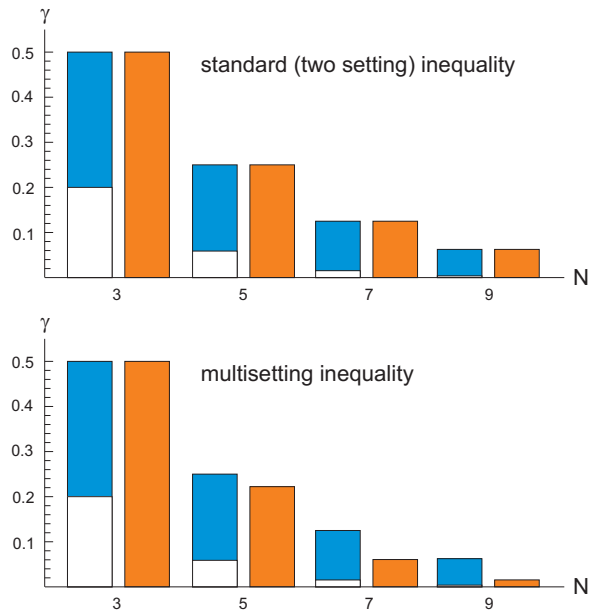}
\end{center}
\caption{Entanglement and violation of the Bell inequalities.
The state $\Upsilon | \mathrm{GHZ}_N \rangle \langle \mathrm{GHZ}_N | + (1-\Upsilon) \rho_{\mathrm{noise}}$
violates corresponding Bell inequality  for values of $\Upsilon$ above the bars.
If the value of $\Upsilon$ lies within blue/orange piece of the bar
the state is entangled, but does not violate the Bell inequality
and therefore does not allow communication complexity reduction.
For each number of qubits, $N$, we present the results
for the white noise admixture (left bar -- blue) and the colored noise $| 0 \rangle \langle 0 |^{\otimes N}$ admixture (right bar -- orange).
For the white noise more settings do not lower the critical admixture.
For the colored noise the critical admixture is lowered.}
\label{f-multi}
\end{figure}

In order to investigate the direct communication-reducing properties of the state we
employ condition~(\ref{ZBCOND}). Depending on the choice of the
observation plane, the left-hand side of Eq.~(\ref{ZBCOND}) reads:
$2^{N-1} \Upsilon^2$ for the $xy$ plane; $(1-\Upsilon)^2 +
\Upsilon^2$ for the $xz$ plane; and $(1-\Upsilon)^2$ for the $yz$
plane. There is no other plane in which the values would be higher,
as the correlation tensor is in its generalized Schmidt form \cite{GEN_SCHMIDT1,GEN_SCHMIDT2}.
The sum over the settings in the $xy$ plane is greater than the sum
over the $xz$ plane for $\Upsilon
> 1/(1+\sqrt{2^{N-1}-1})$. Thus, for the state $\rho(\Upsilon)$ the
left-hand side of (\ref{ZBCOND}) is given by
\begin{equation}
\mathcal{C} = \Bigg\{
\begin{array}{rcc}
2^{N-1} \Upsilon^2 & \textrm{ for } & \Upsilon \leq \frac{1}{1+\sqrt{2^{N-1}-1}}, \\
(1-\Upsilon)^2 + \Upsilon^2 & \textrm{ for } & \Upsilon >
\frac{1}{1+\sqrt{2^{N-1}-1}}.
\end{array}
\end{equation}
Therefore, there exists a local realistic model for the correlations
obtained in any two-setting correlation Bell experiment 
if $\Upsilon \le \Upsilon_{\mathrm{lr}} = 1/\sqrt{2^{N-1}}$,
which is the same critical value as for the state with white noise
(in full analogy to the case of Bell state).
Finally, for $\Upsilon \in
(0,\Upsilon_{\mathrm{lr}}]$ entangled state
$\rho(\Upsilon)$ admits local realistic description for such Bell experiments.
Since the dimension of the system is $d = 2^N$,
the range of Werner gap 
scales polynomially as $d^{-\frac{1}{2}}$.
This is exponentially better than in \cite{ALMEIDA}
where the range of Werner gap scales logarithmically as $\log(d)/d$.
However, the model of \cite{ALMEIDA}
works for arbitrary number of settings
whereas here we have studied only two-setting Bell inequalities for correlation functions.
Already for the multisetting Bell inequalities for correlation functions
\cite{LPZB} the range of the corresponding Werner gap is smaller. 
The left-hand side of (\ref{ZBCOND1}) is given by $\mathcal{D} = 2^{N-1} \Upsilon^2$  for $N=3$, and 
$\mathcal{D}=(1-\Upsilon)^2 + 2^{N-2} \Upsilon^2$ for $N \geq 5$.
This is illustrated in Fig. \ref{f-multi}, where we show the critical entanglement admixtures below which the state satisfies the inequalities.
Note that in this case the fact that the condition is satisfied does not guarantee the existence of the local realistic
model, because this set of inequalities is not necessary and sufficient for the existence of such a model \cite{LPZB}.
We also checked that for the colored noise, the inequalities with continuous settings \cite{ZUK} do not improve the critical admixture
for violation of local realism over the multisetting inequalities \cite{LPZB}, except for $N=3$.

\subsection{Generalized GHZ states}

\subsubsection{Colored noise}

We give an explicit example of a noisy separable state
for which even an infinitesimal small admixture of an infinitesimal weak entangled state 
results in a non-separable state. Consider generalized GHZ state \cite{GEN_GHZ1,GEN_GHZ2,LPZB}:
\begin{equation}
| \mathrm{GHZ(\alpha)} \rangle = \cos\alpha | 0 \dots 0 \rangle + \sin\alpha | 1 \dots 1 \rangle.
\end{equation}
It has the following non-vanishing components
of the correlation tensor:
\begin{eqnarray}
T_{\underbrace{y...y}_{2k}x....x} & = & (-1)^{k} \sin 2 \alpha, \quad k = 0,1,...,\lfloor \tfrac{N-1}{2} \rfloor \nonumber \\
T_{\underbrace{z...z}_{k}0...0} & = & 
\Bigg\{ \begin{array}{ll}
1 & \textrm{ for } k \textrm{ even}, \\
\cos 2 \alpha & \textrm{ for } k \textrm{ odd}. \\
\end{array}
\end{eqnarray}
and similarly for all permutations of indices.

We mix this state with a colored noise 
$\rho_{\mathrm{noise}} = \ket{0} \bra{0}^{\otimes N}$.
If the number of qubits is {\it even},
this state is entangled and violates
Bell inequalities
for both infinitesimal $\alpha$ and $\Upsilon$.
This follows from the fact that the state has perfect correlations
$T_{z...z} = 1$  and additional correlations in complementary directions, e.g., $T_{x...x} = \Upsilon \sin 2 \alpha$.
Therefore, summing up the correlations in the $xz$ plane, using the multisetting condition (\ref{ZBCOND1}) proves the violation.

For {\it odd} number of qubits, the range of Werner gap
again scales (independently of $\alpha$)
polynomially with $d$.
First, we show that the mixed state is entangled already for infinitesimal $\alpha$ and $\Upsilon$,
irrespectively of the number of qubits.
To this aim take two non-vanishing metric elements $G_{zz0...0}$ and $G_{x...x}$ to be equal to $1$.
For this choice, the right-hand side of (\ref{CRITERION})
equals $R = 1 + \Upsilon^2 \sin^2 2 \alpha$.
The left-hand side of the condition now reads
$L = \max (\Upsilon \sin 2 \alpha T_x^{(1)} ... T_x^{(N)} + T_z^{(1)} T_z^{(2)})$,
where we maximize over the choice of local tensors (vectors) $\hat T^{(n)}$.
We set $T_x^{(n)}$ to the maximal value of $1$ for all the parties $n > 2$,
and write the tensor elements for the remaining two parties in polar coordinates,
$L = \max_{\theta_1, \theta_2} (\Upsilon \sin 2 \alpha \sin \theta_1 \sin \theta_2 + \cos \theta_1 \cos \theta_2)$.
Since $\Upsilon \sin 2 \alpha \le 1$, we have $L \le \max_{\theta_1, \theta_2} \cos(\theta_1 - \theta_2) \le 1$.
The maximum is equal to $1$ and for all allowed $\alpha > 0$ and $\Upsilon > 0$
it is smaller than the right-hand side. The state is entangled.

For violation of Bell inequalities,
consider summation over $xy$ plane in the necessary and sufficient condition (\ref{ZBCOND1}).
For the present state, it involves $\sum_{k=0}^{(N-1)/2} {N \choose 2k} = 2^{N-1}$ terms,
each equal to $\Upsilon^2 \sin^2 2 \alpha$,
and gives the critical value of $\Upsilon_{\mathrm{lr}} = \frac{\sqrt{2}}{\sin2\alpha} 2^{-N/2}$.
Therefore the gap $|\Upsilon_{\mathrm{lr}}-\Upsilon_{\mathrm{ent}}|$ scales polynomially 
with dimension as $1/\sqrt{d}$ for $d=2^N$.

\subsubsection{Local depolarization}

The non-vanishing correlation tensor elements,
after local depolarizing channels are applied
to the generalized GHZ state, read
\begin{eqnarray}
T_{\underbrace{y...y}_{2 k}x....x} & = & (-1)^{k} p^N \sin 2 \alpha, \quad k = 0,1,...,\lfloor \tfrac{N-1}{2} \rfloor \nonumber \\
T_{\underbrace{z...z}_{k}0...0} & = & 
\Bigg\{ \begin{array}{ll}
p^k & \textrm{ for } k \textrm{ even}, \\
p^k \cos 2 \alpha  & \textrm{ for } k \textrm{ odd}.
\end{array}
\end{eqnarray}

To show the Werner gap for $N \to \infty$,
we first prove that the state is entangled for all $p> \tfrac{1}{2}$.
Choose the following non-zero elements of the metric: $G_{j_1...j_N} = 1$
for $j_n = 1,2$, $G_{z...z} = 1$ for $N$ even
and $G_{z...z0} = 1$ for odd $N$.
The right-hand side of the entanglement condition (\ref{CRITERION})
is $R = p^{2(N-1+N_2)} + 2^{N-1} p^{2(N-1+N_2)} \sin^2 2 \alpha$,
where $N_2 = N \mod 2$ encodes the cases of odd and even $N$.
The left-hand side is maximized if all local tensors are the same
and along the $z$ axes and have the value of $L = p^{N-1+N_2}$.
Therefore, the state is entangled if $1 < p^{N-1+N_2} + 2^{N-1} p^{N-1+N_2} \sin^2 2 \alpha$.
To unify the cases of odd and even $N$
we bound the right-hand side from below using $p^N \le p^{N-1+N_2}$,
and obtain the sufficient condition for entanglement.
The corresponding critical value is
\begin{equation}
p_{\mathrm{ent}} = (1 + 2^{N-1} \sin^2 2 \alpha)^{-\frac{1}{N}} \to \tfrac{1}{2},
\end{equation}
where the limit is for $N \to \infty$.
Arrows in following formulae
always denote this limit.

The multisetting Bell inequalities give better results than standard inequalities for this state.
Consider the violation condition (\ref{ZBCOND1})
in which the last index of the correlation tensor takes on the values $\{y,z\}$,
whereas indices of other parties are either $\{x,y\}$, 
if the last index is $y$, or $z$, if the last index is $z$.
(Note that we explicitly make use here of the advantage
of the multisetting condition over the two-setting condition).
The value of parameter $\mathcal{D}$ is at least
(it is higher for $N$ even) equal to
$p^{2N}(\cos^2 2 \alpha + 2^{N-2} \sin^2 2 \alpha)$,
where $2^{N-2}$ gives the number of non-zero
correlation tensor elements
with $y$ index at last position.
Therefore, the critical parameter is
\begin{equation}
p_{\mathrm{lr}} = (\cos^2 2 \alpha + 2^{N-2} \sin^2 2 \alpha)^{-\tfrac{1}{2N}} \to \tfrac{1}{\sqrt{2}}.
\end{equation}
The critical values of $p$ decrease with the number of qubits
showing that many party generalized GHZ states are more and more
robust against this type of noise.
Finally, for $N \to \infty$ there is a Werner gap of $p \in (\frac{1}{2},\frac{1}{\sqrt{2}})$
for which the entangled state does not improve related communication complexity tasks.

\subsubsection{Dephasing}

Similar results hold for other local noises.
After dephasing in the local $z$-bases,
the correlation tensor of the generalized GHZ state reads
\begin{eqnarray}
T_{\underbrace{y...y}_{2 k}x....x} & = & (-1)^{k} \lambda^{\frac{N}{2}} \sin 2 \alpha, \quad k = 0,1,...,\lfloor \tfrac{N-1}{2} \rfloor \nonumber \\
T_{\underbrace{z...z}_{k}0...0} & = & 
\Bigg\{ \begin{array}{ll}
1 & \textrm{ for } k \textrm{ even}, \\
\cos 2 \alpha  & \textrm{ for } k \textrm{ odd}.
\end{array}
\end{eqnarray}

All entangled generalized GHZ states
are still entangled after the local dephasing.
For a proof, it is sufficient to choose $G_{zz0..0} = G_{x..xx} = 1$.
For this choice, the right-hand side of (\ref{CRITERION})
reads $R = 1 + \lambda^N \sin^2 2 \alpha$,
whereas for the left-hand side we have
$L = \max\left( T_z^{(1)} T_z^{(2)} + T_x^{(1)} T_x^{(2)} \lambda^{N/2} \sin2\alpha \right)
\le T_z^{(1)} T_z^{(2)} + T_x^{(1)} T_x^{(2)} \le 1$,
which follows from $\lambda^{N/2} \sin2\alpha \le 1$
and writing the components of the local tensors
in polar coordinates.
We also assumed that the local Bloch vectors
of all the parties except first and second
are along the $x$ axis, which is optimal.
Therefore, the state is entangled
for all $\alpha > 0$ and $\lambda >0$,
independently of the number of qubits.

Since dephasing leaves the correlations
in specific directions unchanged,
violation of Bell inequality for the generalized GHZ state is very robust against this type of noise.
We show that it actually is state independent,
i.e. violation is observed for all $\alpha > 0$
and only depends on the degree of dephasing if $N$ is odd.
Consider the mutisetting condition,
in which as before the last index takes values $\{y,z\}$
and the remaining indices are either $\{x,y\}$,
if the last index is $y$, or $z$, if the last index is $z$.
If $N$ is even, after dephasing the state still contains
perfect correlations in the $z$ directions
and some other correlations in the $xy$ plane,
and violates the inequalities for all $\alpha > 0$ and $\lambda >0$.
For the case of odd number of qubits,
the condition reads
$\mathcal{D} = \cos^2 2 \alpha + 2^{N-2} \lambda^N \sin^2 2 \alpha$,
and the violation is observed
as soon as
\begin{equation}
\sin^2\alpha > 0 \textrm{ and } \lambda > 2^{\frac{2}{N}-1} \to \tfrac{1}{2}.
\end{equation}
Therefore, violation only depends on the degree of dephasing
in the case of odd $N$, and again there is a finite Werner gap of
$\lambda \in (0,\frac{1}{2})$ in the limit $N \to \infty$.

\subsubsection{Amplitude damping}

Finally, we consider independent local amplitude damping channels.
The elements of the decohered generalized GHZ states
are the following:
\begin{eqnarray}
T_{\underbrace{y...y}_{2 k}x....x} & = & (-1)^{k} \gamma^{\frac{N}{2}} \sin 2 \alpha, \quad k = 0,1,...,\lfloor \tfrac{N-1}{2} \rfloor \nonumber \\
T_{\underbrace{z...z}_{k}0...0} & = & 
\Bigg\{ \begin{array}{ll}
\cos^2\alpha+\bar \gamma^k \sin^2 \alpha & \textrm{ for } k \textrm{ even}, \\
\cos^2\alpha-\bar \gamma^k \sin^2 \alpha  & \textrm{ for } k \textrm{ odd},
\end{array}
\end{eqnarray}
where $\bar \gamma = 2 \gamma -1$.
To prove the Werner gap consider the metric
with non-vanishing elements $G_{j_1...j_N} = 1$
with $j_n=1,2$.
The right-hand side of the entanglement criterion
reads $R = 2^{N-1}\gamma^N \sin^2 2 \alpha$.
The left-hand side is $L = \gamma^{N/2} \sin 2\alpha \max[T_x^{(1)} \dots T_x^{(N)} - T_y^{(1)}T_y^{(2)}T_x^{(3)} \dots T_x^{(N)} - \dots ]$,
with the maximum taken over local tensors with components from one plane.
We write the elements of individual tensors in polar coordinates,
i.e. $T_x^{(n)} = \cos \theta_n$ and $T_y^{(n)} = \sin \theta_n$,
and recognize that expression in the bracket is now given by
$\cos(\theta_1 + \ldots + \theta_N)$. Therefore, $L = \gamma^{N/2} \sin 2\alpha$,
which translates into critical parameter for entanglement
\begin{equation}
\gamma_{\mathrm{ent}} = \frac{1}{4}\left( \frac{2}{\sin 2 \alpha}\right)^{\frac{2}{N}} \to \frac{1}{4}.
\end{equation}

The violation of both many-setting and two-setting inequalities
is the same for higher number of qubits.
The two-setting condition reveals the critical parameter
\begin{equation}
\gamma_{\mathrm{lr}} = \frac{1}{2} \left(\frac{2}{\sin 2 \alpha}\right)^{\frac{1}{N}} \to \frac{1}{2}.
\end{equation}
For large $N$, the states for practically all $\alpha > 0$
violate the inequalities if $\Upsilon_{\mathrm{lr}} > 1/2$ and there is a finite Werner gap 
of $\gamma \in (\frac{1}{4},\frac{1}{2})$ in the limit $N \to \infty$.

\subsection{$W$ state}

In this section we study the $W$ state,
and we shall emphasize the properties distinguishing it form the class of generalized GHZ states.
The $W$ state involves a single excitation
delocalized over all the qubits:
\begin{equation}
| W \rangle = \tfrac{1}{\sqrt{N}} \left( | 1 0  \ldots  0 \rangle + | 0 1 \ldots  0 \rangle + \ldots +| 0 0  \ldots  1 \rangle \right).
\end{equation}
It is permutationally invariant,
i.e. any permutation of particles leaves the state unchanged.
Therefore, to describe its correlation tensor
it is sufficient to present just three elements.
All other non-vanishing elements
have indices being permutations of the indices
of the following ones:
\begin{eqnarray}
T_{\underbrace{z...z}_{k} 0...0} & = & 1 - \frac{2 k}{N},\nonumber \\
T_{yyz...z0...0} = T_{xxz...z0...0} & = & \frac{2}{N}.
\end{eqnarray}

\subsubsection{White noise}

Consider $W$ state mixed with the white noise:
\begin{equation}
\rho = \Upsilon | W \rangle \langle W | + (1-\Upsilon) \frac{1}{2^N}\openone.
\end{equation}
Contrary to the case of mixed GHZ state, 
this mixed state gives rise to a Werner gap
in the limit of $N \to \infty$.

To prove entanglement of this state
consider metric with non-vanishing elements
$G_{j_1...j_N} = 1$ where $j_n = \{x,z\}$.
With this choice,
the right-hand side of condition (\ref{CRITERION})
reads $R = \Upsilon^2(1 + {N \choose 2} \frac{4}{N^2}) = \Upsilon^2(3-\tfrac{2}{N})$.
The left-hand side is maximized,
if all the local tensors are along the $\pm z$ axis
and equals $L = \Upsilon$, which we have verified numerically.
Therefore, the critical parameter for entanglement reads
\begin{equation}
\Upsilon_{\mathrm{ent}}  = \frac{1}{3-\tfrac{2}{N}} \to \frac{1}{3}.
\end{equation}

The multisetting inequalities are violated
as soon as entanglement admixture is above the critical value \cite{LPZB}:
\begin{equation}
\Upsilon_{\mathrm{lr}} = \sqrt{\tfrac{1}{3-\frac{2}{N}}} \to \frac{1}{\sqrt{3}}.
\end{equation}
Note that the same correlations enter
both the Bell inequalities and the entanglement criterion,
showing that this simple application of the criterion
is at least quadratically better in revealing entanglement than the Bell inequalities, i.e. $\Upsilon_{\mathrm{ent}} \le \Upsilon_{\mathrm{lr}}^2$.
This is a general feature present in all our examples.

\subsubsection{Colored noise}

Similar conclusion for comparison with the GHZ states follows
in the case of colored noise admixture to the $W$ state:
\begin{equation}
\rho = \Upsilon | W \rangle \langle W | + (1 - \Upsilon) |0...0 \rangle \langle 0...0 |,
\end{equation}
where the colored noise introduces correlations in the local $z$ directions.

This state is entangled for all $\Upsilon > 0$.
A simple way to see this is to use the fact that if a subsystem is entangled, then the whole system is also entangled.
The definition of the $W$ state
leads to the following form of the reduced density operator for any two qubits
\begin{equation}
\Upsilon' | \psi^+ \rangle \langle \psi^+ | + (1-\Upsilon') |00 \rangle \langle 00|,
\end{equation}
with $\Upsilon' = \frac{2}{N} \Upsilon$ and $| \psi^+ \rangle = \frac{1}{\sqrt{2}}(\ket{01} + \ket{10})$. 
This state is entangled (has negative partial transposition \cite{PERES_PPT,HORODECKI_PPT})
for all $\Upsilon' > 0$.
Therefore the global state is entangled for all $\Upsilon_{\mathrm{ent}} > 0$ and any finite $N$.

If one chooses settings from the $xz$ plane in the multisetting Bell inequality,
the violation conditions reveal the critical value of the admixture parameter
as given by
\begin{equation}
\Upsilon_{\mathrm{lr}} = \frac{2}{3-\frac{1}{N}} \to \frac{2}{3}.
\end{equation}
In the limit of $N \to \infty$ we find the Werner gap of $\Upsilon \in (0,\frac{2}{3})$.
This is the biggest gap among all the states studied here.

\subsubsection{Local depolarization}

We shall show that $W$ state is very fragile with respect to this type of decoherence,
in contrast to the GHZ state.
After local depolarization the elements of the correlation tensor of the decohered $W$ state read
\begin{eqnarray}
T_{\underbrace{z...z}_{k} 0...0} & = & p^k \left(1 - \frac{2 k}{N} \right), \nonumber \\
T_{yy\underbrace{z...z}_{k}0...0} = T_{xx\underbrace{z...z}_{k}0...0} & = & p^{k+2} \frac{2}{N}.
\end{eqnarray}

To prove entanglement of this state, consider non-zero metric elements
$G_{j_1...j_N} = 1/p^N$ with $j_n = \{x,z\}$.
For this choice the right-hand side of the criterion equals $R = p^N(1 + {N \choose 2} \frac{4}{N^2})$,
whereas the maximum of the left-hand side is $1$,
which we have verified numerically.
Therefore, the state is entangled above the critical value of
\begin{equation}
p_{\mathrm{ent}} = \left( \frac{1}{3-\frac{2}{N}} \right)^{\frac{1}{N}} \to 1.
\end{equation}

The multisetting Bell inequalities are violated
as soon as $\mathcal{D} = T_{z...z}^2 + {N \choose 2} T_{xxz...z}^2 > 1$.
This gives the critical parameter
\begin{equation}
p_{\mathrm{lr}} = \left( \frac{1}{3-\frac{2}{N}} \right)^{\frac{1}{2 N}} \to 1.
\end{equation}
which rapidly increases with $N$ and already for five qubits requires $p>0.9$.
Since $p_{\mathrm{ent}} = p_{\mathrm{lr}}^2$, there is a Werner gap for all finite $N$,
and in the limit both parameters tend to the same value.
Of course, a smarter choice of the metric in the entanglement condition
could prove that even in the limit there is a finite Werner gap.

\subsubsection{Dephasing}

The $W$ state is extremely robust against dephasing,
as it leaves the perfect correlations unchanged.
After dephasing the $W$ state is transformed to
\begin{eqnarray}
T_{\underbrace{z...z}_{k} 0...0} & = & 1 - \frac{2 k}{N}, \nonumber \\
T_{yy\underbrace{z...z}_{k}0...0} = T_{xx\underbrace{z...z}_{k}0...0} & = & \lambda \frac{2}{N}.
\end{eqnarray}

The dephased state violates Bell inequality
(and therefore is entangled) for all $N$
and all non-trivial dephasing channels.
Consider correlations in the $xz$ plane and multisetting condition.
The value of parameter $\mathcal{D} = 1 + 2 \lambda^2(1-\tfrac{1}{N})$
exceeds unity for all $\lambda > 0$.
Note that this is true also in the limit $N \to \infty$.

\subsubsection{Amplitude damping}

The $W$ state after this type of decoherence reads
\begin{eqnarray}
T_{\underbrace{z...z}_{k}0...0} & = & 1 - \frac{2 k}{N} \gamma, \nonumber \\
T_{yy\underbrace{z...z}_{k}0...0} = T_{xx\underbrace{z...z}_{k}0...0} & = & \frac{2}{N}\gamma.
\end{eqnarray}

To prove the Werner gap,
consider non-vanishing metric elements $G_{j_1...j_N} = 1$ for $j_n = \{x,z\}$.
The right hand side of the criterion is $R = (1 - 2 \gamma)^2 + {N \choose 2} \frac{4}{N^2}$,
and the maximum of the left-hand side $L \le 1 - 2 \gamma$
is attained for all local vectors along $z$ directions.
Therefore, the state is entangled above the critical value
\begin{equation}
\gamma_{\mathrm{ent}} = \frac{1}{3 - \frac{1}{N}} \to \frac{1}{3}.
\end{equation}

We check violation of Bell inequalities
using the condition with many settings in the $xz$ plane.
The expression reads $\mathcal{D} = R$ and exceeds unity
for all values above 
\begin{equation}
\gamma_{\mathrm{lr}} = \frac{2}{3 - \frac{1}{N}} \to \frac{2}{3}.
\end{equation}
The Werner gap is present also in the limit $N \to \infty$,
just as for the GHZ state.

\section{Summary}

\begin{table*}
\begin{tabular}{l l l l l l}
\hline \hline
                     &     white noise $\qquad$     &            colored noise   $\quad$                 &  local depolarization $\quad$ & dephasing $\qquad$ & amplitude damping \\ \hline \hline
Gen. GHZ $\quad$  & $\zeta_{\mathrm{ent}} \to 1/2$ & $\zeta_{\mathrm{ent}} \to 0$ & $\zeta_{\mathrm{ent}} \to 1/2$ & $\zeta_{\mathrm{ent}} \to 0$ & $\zeta_{\mathrm{ent}} \to 1/4$ \\
                     & $\zeta_{\mathrm{lr}} \to 1/\sqrt{2}$ & $\zeta_{\mathrm{lr}} \to 1/\sqrt{2}$ & $\zeta_{\mathrm{lr}} \to 1/\sqrt{2}$ & $\zeta_{\mathrm{lr}} \to 1/2$ & $\zeta_{\mathrm{lr}} \to 1/2$ \\ \hline 
W & $\zeta_{\mathrm{ent}} \to 1$ & $\zeta_{\mathrm{ent}} \to 0$ & $\zeta_{\mathrm{ent}} \to 1$ & $\zeta_{\mathrm{ent}} \to 0$ & $\zeta_{\mathrm{ent}} \to 1/3$ \\
    & $\zeta_{\mathrm{lr}} \to 1$ & $\zeta_{\mathrm{lr}} \to 1$ & $\zeta_{\mathrm{lr}} \to 1$ & $\zeta_{\mathrm{lr}} \to 0$ & $\zeta_{\mathrm{lr}} \to 2/3$  \\ \hline \hline
\end{tabular}
\caption{Summary of the results. The results for different initial states are presented in rows.
Noisy channels applied to them are presented in columns.
The strength of the noises is characterized by a single parameter $\zeta$
($\zeta = 1$ corresponds to no noise, $\zeta = 0$ describes the strongest noise
which immediately destroys entanglement).
To unify presentation in this table we represent all the parameters with $\zeta$. 
In the main text the parameter characterizing local depolarization is denoted by $p$, dephasing by $\lambda$, amplitude damping by $\gamma$,
and admixture of white or colored noise by $\Upsilon$.
We present the critical parameter $\zeta_{\mathrm{ent}}$,
above which the resulting state is entangled,
and $\zeta_{\mathrm{lr}}$,
below which the state satisfies classes of Bell inequalities,
in the limit of large number of qubits $N \to \infty$.
Additionally, to compare on equal footing the critical parameters for different types of noises,
they are all calculated here per particle
in a sense that the values related to white and colored noise
are $N$th roots of the values of the main text.
In all the cases, $\zeta_{\mathrm{ent}}$ is at most a square of $\zeta_{\mathrm{lr}}$.}
\label{TAB_SUMMARY}
\end{table*}

Using the entanglement criterion \cite{BADZIAG} we have found families of entangled states 
which satisfy specific classes of Bell inequalities.
We summarize our findings in Table \ref{TAB_SUMMARY}.
Generally speaking, a simple application of the entanglement criterion gives
at least quadratically better critical parameters
than the ones obtained using the Bell inequalities.
Therefore, we found entangled states satisfying the Bell inequalities in all studies cases.
Moreover, we gave examples in which even in the limit of large
number of qubits there is a finite gap between critical parameter for entanglement
and critical parameter for Bell violation.
We found that maximally entangled mixed states of two qubits
give rise to the highest discrepancy between the critical parameters.
It would be interesting to investigate if this also holds for higher number of qubits.
Our results are a further
step towards full classification of entangled states into 
those which do and do not admit local realistic explanation.

\section{Acknowledgments}

We thank Johannes Kofler and Ravishankar Ramanathan for discussions.
W.L. and M.\.Z. performed this work at National Quantum Information Centre of Gda\'nsk.
W.L. is supported by the Foundation for Polish Science (KOLUMB program). M. \.Z. acknowledges EU program SCALA.
We acknowledge support of the Austrian Science Foundation FWF Project No. P19570-N16,
SFB and the Doctoral Program CoQuS,the $6$th EU program QAP (Qubit Applications) Contract No. 015848,
and the National Research Foundation \& Ministry of Education in Singapore.
The collaboration is a part of an \"OAD/MNiSW program.

\end{document}